\documentstyle{elsartwb}

\input epsf.sty

\begin{document}

\begin{frontmatter}
\title{\bf Boost-invariant particle production in transport equations$^*$ } 
\thanks{Research supported in part by the Polish State Committee for
        Scientific Research, grant 2 P03B 09419 }
\author{Katarzyna Bajan, Wojciech Florkowski}
\address{The H. Niewodnicza\'nski Institute of Nuclear Physics, \\
        ul. Radzikowskiego 152,  PL-31342 Krak\'ow, Poland}

\begin{abstract}
The boost-invariant tunneling of particles along the hyperbolas of
constant invariant time $\tau=\sqrt{t^2-z^2}$ is included in the
transport equations describing formation of the quark-gluon plasma in
strong color fields. The non-trivial solutions of the transport
equations exist if the boost-invariant distance between the tunneling
particles, measured in the quasirapidity space $\eta=1/2 \ln
((t+z)/(t-z))$, is confined to a finite interval $\Delta \eta$.  For
realistic values of $\Delta \eta$ the
solutions of the transport equations show similar characteristics to
those found in the standard approach, where the tunneling takes place
at constant time $t$.  In the limit $\Delta \eta \rightarrow
\infty$, the initial color fields decay instantaneously.

\end{abstract}
\end{frontmatter}
\vspace{-7mm} PACS: 25.75.-q, 05.20.Dd, 24.85.+p

\section{Introduction}

In this paper we solve transport equations describing production of the
quark-gluon plasma in strong color fields. A novel feature of our approach
is the implementation of the boost-invariant tunneling along the hyperbolas
of constant invariant time $\tau =\sqrt{t^{2}-z^{2}}$ \ into the framework
of the relativistic kinetic theory. In this way we generalize the previous
results of Refs. \cite{BCapp,BCDFosc,DFcft,DFhq}.

In the standard WKB description of the tunneling process \cite{cnn,gm}, the
particles tunnel at fixed time $t.$ They emerge from the vacuum at a certain
distance from each other, and their longitudinal momenta vanish (in the
center-of-mass frame of a pair). In Refs. \cite{BCapp,BCDFosc,DFcft,DFhq,km}%
, the WKB results were used to fix the longitudinal-momentum dependence of
the production rates of quarks and gluons. The effect of the finite distance
appearing between the tunneling particles is more difficult to include. One
assumes usually that this distance is small, and the non-local character of
the tunneling process is neglected. This approximation is not always valid.
For large transverse mass, the distance between the tunneling particles may
be quite substantial and should be taken into account.

The main difficulty connected with the non-local features of the tunneling
concerns the causal properties of sequential decays. In this case one cannot
define which pair is produced earlier or later, which leads to ambiguity in
the determination of the decay probabilities. Nevertheless, at very high
energies a solution to this problem exists \cite{BCDFtun}. Assuming that the
pairs are produced at fixed invariant time $\tau =\sqrt{t^{2}-z^{2}}$, one
introduces a Lorentz-invariant sequence of pair production, and the
requirements of causality can be naturally fulfilled. The tunneling of
particles along the hyperbolas of constant invariant time leads to finite
longitudinal momenta of the created particles. Thus, the production rates in
the kinetic equations should include also this extra effect.

The boost-invariant description of the tunneling process was the main
ingredient of the simulation program used to describe the space-time
evolution of the color-flux tubes \cite{DFsim,Dyrek}. In particular, this
program was applied in the investigations of intermittency \cite
{BCDFPint} and soft photon emission \cite{CFphot}. In the present paper
the boost-invariant tunneling along the hyperbolas of constant invariant
time is included in the kinetic equations. In contrast to the simulation
studies \cite{BCDFPint,CFphot}, which were restricted to the case of the
elementary color fields, we now deal with stronger fields. Similarities and
differences between the present approach and Refs. \cite{BCapp,BCDFosc} are
discussed in detail.
 
\section{Pair production in strong chromoelectric fields}

\subsection{Boost-invariant tunneling}

A semi-classical boost-invariant description of pair production in
chromoelectric fields was introduced in Ref. \cite{BCDFtun}. In this
approach the tunneling particles move along the hyperbola of constant
invariant time
\begin{equation}
\tau =\sqrt{t^{2}-z^{2}}.  \label{tau}
\end{equation}
If the virtual particles start their motion at $z=0$, see Fig. 1, the
end points of the tunneling trajectory may be determined from the
energy-momentum conservation laws \
\begin{equation}
E_{f}=\sqrt{m_{\perp }^{2}+p_{f\parallel }^{2}}=F\ z_{f},\qquad
p_{f\parallel }=F\left( t_{f}-\tau \right) .  \label{eppar}
\end{equation}
Here $E_{f}$ and $p_{f\parallel }$ are the energy and the longitudinal
momentum of the particle at the space-time point where it emerges from the
vacuum, $m_{\perp}$ is the transverse mass, and $F$ is a constant
force acting on the particle (in this Section we shall concentrate our 
discussion on the case $F>0$ only). Using the rapidity variable $y$ we may write 
\begin{equation}
E_{f}=m_{\perp }\cosh y_{f}\ ,\qquad p_{f\parallel }=m_{\perp }\sinh y_{f}\ .
\label{yf}
\end{equation}
In the analogous way we define the quasirapidity variable $\eta $, which
gives 
\begin{equation}
t_{f}=\tau \cosh \eta _{f}\ ,\qquad z_{f}=\tau \sinh \eta _{f}.
\label{etaf}
\end{equation}
Equations (\ref{eppar}), (\ref{yf}) and (\ref{etaf})\ yield \cite{BCDFtun} 
\begin{equation}
\sinh y_{f}=\frac{m_{\perp }}{2F\tau},\qquad \eta _{f}=2\ y_{f}.  
\label{sinhyf}
\end{equation}

\begin{figure}[b]
\epsfysize=9cm
\par
\begin{center}
\mbox{\epsfbox{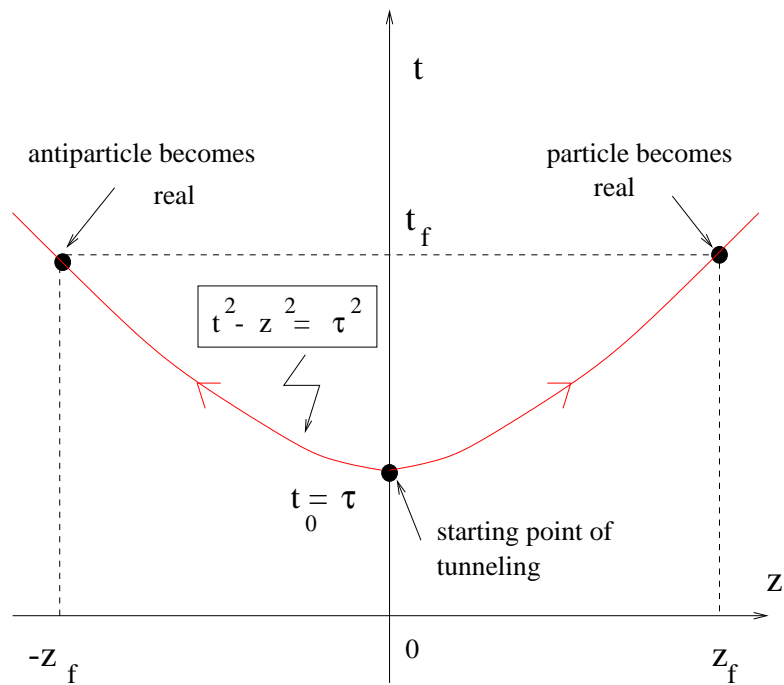}}
\end{center}
\caption{Boost-invariant tunneling of particles along the 
trajectory of constant invariant time $\tau=\sqrt{t^2-z^2}$}
\label{fig1}
\end{figure}

An interesting feature of Eq. (\ref{sinhyf}) is that the kinematical
quantities characterizing the tunneling particles depend on $\tau $, i.e.,
pairs with different momenta are created at different invariant times. This
is an effect of the boundary conditions which describe expansion of the
system. For very large $\tau $, when the boundary of the field is far away
from the center of the system we recover the standard results: $p_{f \parallel
}=0,\ z_{f}=m_{\perp }/F$ and $t_{f}=\tau $.

Although the boundary conditions change the kinematics of the tunneling
process, the probability of tunneling per unit volume of space-time does not
change \cite{BCDFtun}. Thus, we may use the well established formula for the
production rate \cite{cnn,gm,schwinger,gi}
\begin{equation}
\frac{dN}{d^{4}x\ d^{2}p_{\perp }}=\frac{F}{4\pi ^{3}}\left| \ln \left( 1\mp
\exp \left( -\frac{\pi m_{\perp }^{2}}{F}\right) \right) \right| .
\label{rate0}
\end{equation}
Here the minus sign is appropriate for fermions (in our case for quarks and
antiquarks) and the plus sign should be used for bosons (in our case for
gluons). Eq. (\ref{rate0})\ gives the rate integrated over the longitudinal
momentum. The $p_{\parallel }$-dependence may be taken into account by the
following modification of (\ref{rate0})\ 
\begin{equation}
\frac{dN}{d\Gamma }\equiv p^{0}\frac{dN}{d^{4}x\ d^{3}p}=\frac{F}{4\pi ^{3}}%
\left| \ln \left( 1\mp \exp \left( -\frac{\pi m_{\perp }^{2}}{F}\right)
\right) \right| \delta \left( y-\eta +\frac{1}{2}\eta _{f}\right) .
\label{rate}
\end{equation}
The production rate (\ref{rate})\ is boost invariant and reduces to the
previous formula\ if divided by $p^{0}$ and integrated over $p_{\parallel }$.
Moreover, the rapidity of the particles which are produced at $\eta =\eta
_{f}$ is simply $y_{f}=\frac{1}{2}\eta _{f}$, the result required by the
condition of the boost-invariant tunneling discussed above. In the limit of
large invariant times or large color fields we find $\eta
_{f}\longrightarrow 0$, hence the standard formula for tunneling is
recovered: the longitudinal momenta of particles tunneling at $z=0$ are zero.

\subsection{Boost-invariant variables $w$ and $v$}

In the next Sections we shall use the boost-invariant variables introduced 
in Refs. \cite{BCprd,BCzfc}
\begin{equation}
u=\tau ^{2}=t^{2}-z^{2},\quad w=tp_{\Vert }-zE,\quad \mathbf{p}_{\bot },
\label{binvv1}
\end{equation}
and also 
\begin{equation}
v=Et-p_{\Vert }\ z=\sqrt{w^{2}+m_{\perp }^{2}u}.  \label{binvv2}
\end{equation}
From these two equations one can easily find the energy and the longitudinal
momentum of a particle 
\begin{equation}
E=p^{0}=\frac{vt+wz}{u},\quad p_{\Vert }=\frac{wt+vz}{u}.  \label{binvv3}
\end{equation}
The invariant measure in the momentum space is
\begin{equation}
dP = d^2p_\perp {dp_\parallel \over p^0} = d^2p_\perp {dw \over v}.
\label{dP}
\end{equation}
In addition we have 
\begin{equation}
w=\tau m_{\perp }\sinh \left( y-\eta \right) ,\qquad v=\tau m_{\perp }\cosh
\left( y-\eta \right) .  \label{binvv4}
\end{equation}
Equation (\ref{binvv4}) allows us to rewrite the production rate in the
form 
\begin{equation}
\frac{dN}{d\Gamma }
= p^{0}\frac{dN}{d^{4}x\ d^{3}p}=\frac{F}{4\pi ^{3}}\left|
\ln \left( 1\mp \exp \left( -\frac{\pi m_{\perp }^{2}}{F}\right) \right)
\right| \delta \left( w-w_{0}\right) v,  \label{rate1}
\end{equation}
where 
\begin{equation}
w_{0}= \tau m_{\perp }\sinh \left( y_{f}-\eta _{f}\right) =-\tau m_{\perp
}\sinh \left( \frac{\eta _{f}}{2}\right) =-\frac{ m_{\perp }^{2}}
{2F}.  \label{wf}
\end{equation}

\subsection{Finite-size corrections}

Comparing Eqs. (\ref{binvv1}) and (\ref{wf}) we find that the
longitudinal momenta of the tunneling particles at $z=0$ are $-
m_{\perp}^{2}/ (2F \tau)$. Thus, for large invariant times $p_{\Vert }
\rightarrow 0$.  On the other hand, in the limit of small invariant
times $p_{\Vert }$ becomes infinite. As we shall see in more detail in
Section 4, this divergence leads to infinite values of color
currents at $\tau=0$, and to infinite decay rate of the initial 
chromoelectric field. The physical origin of this singularity is
the possibility of the creation of particles at infinite values of
$t$ and $z$, which correspond to small values of the invariant time $\tau$.
In practice, we always deal with finite systems and such tunneling 
cannot take place. In order to take into account such finite-size
corrections we impose an additional condition on the tunneling process,
namely \cite{Dyrek}
\begin{equation}
2 \eta_f = 4 \hbox{Arcsinh} \left( \frac{m_{\perp }}{2F\tau} \right)
< \Delta \eta.
\label{deltaeta}
\end{equation}
Here $\Delta \eta$ is a parameter which determines the space-time
region in quasirapidity, $|\eta| < \Delta \eta$, where the tunneling
is possible. The tunneling particles should fit into this region,
hence the distance between the emerging members of a pair
should be smaller than $\Delta \eta$. 
As a consequence, our final expression for
the production rate is
\footnote{We neglect here the boundary effects:
The tunneling processes can start from a place close to the edge of the
system and may have not enough space to fit into the allowed
region. Such situations should be eliminated by introducing
an additional constraint.}
\begin{eqnarray}
\!\!\!\!\!\!\frac{dN}{d\Gamma }&=&\frac{F}{4\pi ^{3}}
\left|
\ln \left( 1\mp \exp \left( -\frac{\pi m_{\perp }^{2}}{F}\right) \right)
\right| 
\theta\left[2\tau F \sinh \left( {\Delta \eta \over 4} \right)- m_{\perp }
\right]
\delta \left( w-w_{0}\right) v, \nonumber \\ 
& &
\label{rate2}
\end{eqnarray}
where $\theta $ is the step function 
\begin{equation}
\theta (x)=1\quad \hbox{for}\quad x>0,\quad \theta (x)=0\quad \hbox{for}%
\quad x\leq 0.  \label{stepfun}
\end{equation}
We note that condition (\ref{deltaeta}) was also used in the
simulation program \cite{DFsim,Dyrek}. In that case, the allowed
region in the quasirapidity space was determined by the actual size of
a decaying color flux tube. In the present study, the size of $\Delta
\eta$ is suggested by the rapidity range accessible in the
ultra-relativistic heavy-ion collisions at SPS and RHIC.
In the following, we consider three typical values:
$\Delta \eta = 4, 6$ and 8.

\section{Semi-classical kinetic equations for quark-gluon plasma}

In the {\it abelian dominance approximation}, the equations for quarks,
antiquarks and gluons have the form 
\cite{BCDFosc,heinz,egv} 
\begin{equation}
\left( p^{\mu }\partial _{\mu } \pm g{\mbox{\boldmath $\epsilon$}}_{i}\cdot 
{\bf F}^{\mu \nu }p_{\nu }\partial _{\mu }^{p}\right) 
G^\pm_{i}(x,p)=\frac{dN^\pm_{i}}{d\Gamma },  \label{kineq}
\end{equation}
\begin{equation}
\left( p^{\mu }\partial _{\mu }+g{\mbox{\boldmath $\eta$}}_{ij}\cdot 
{\bf F}^{\mu \nu }p_{\nu }\partial _{\mu }^{p}\right) 
\tilde{{G}}_{ij}(x,p)=
\frac{d{\tilde{N}}_{ij}}{d\Gamma },  \label{kineg}
\end{equation}
where $G^+_{i}(x,p)\ $, $G^-_{i}(x,p)$ and $\tilde{{G}}_{ij}(x,p)$
are the phase-space densities of quarks, antiquarks and gluons, respectively. 
Here $g$ is the strong coupling constant, and $i,j=(1,2,3)$ are color
indices. The terms on the left-hand-side describe the free motion of the
particles and the interaction of the particles with the mean field $\mathbf{F%
}_{\mu \nu }$. The terms on the right-hand-side describe production of
quarks and gluons due to the decay of the field. The distribution functions
$G^{\pm}_i$ and $\tilde{{G}}_{ij}$ include the spin degeneracy factors.
In the numerical calculations we neglect the quark masses and assume that 
Eq. (\ref{kineq}) holds for $N_f$=3 flavors. We note that Eqs. (\ref{kineq})
and (\ref{kineg}) do not include the thermalization effects. The latter can
be taken into account in the relaxation time approximation 
(see for example Refs. \cite{bbr,bn,brs}).

The only non-zero components of the tensor ${\bf F}_{\mu \nu }=(F_{\mu
\nu }^{3},F_{\mu \nu }^{8})$ are those corresponding to the chromoelectric
field ${\mbox{\boldmath $\cal E$}}$, which may be written as 
\begin{equation}
{\mbox{\boldmath $\cal E$}}={\bf F}^{30}{=-2}\frac{d\mathbf{h}}{du}=-%
\frac{1}{\tau }\frac{d\mathbf{h}}{d\tau }.  \label{binve}
\end{equation}
Here $\mathbf{h}$ is a function of the variable $u=\tau ^{2}$ only
(note that ${\mbox{\boldmath $\cal E$}}$ is invariant under Lorentz
boosts along the $z$-axis). The quarks couple to the chromoelectric
field ${\mbox{\boldmath $\cal E$}}$ through the charges \cite{huang}
\begin{equation}
{\small \mbox{\boldmath $\epsilon$}}_{1}{\small =}\frac{1}{2}\left( 1,\sqrt{%
\frac{1}{3}}\right) {\small ,\mbox{\boldmath $\epsilon$}}_{2}{\small =}\frac{%
1}{2}\left( -1,\sqrt{\frac{1}{3}}\right) {\small ,\mbox{\boldmath $\epsilon$}%
}_{3}{\small =}\left( 0,-\sqrt{\frac{1}{3}}\right) {\small .}
\label{qcharge}
\end{equation}

The gluons couple to ${\mbox{\boldmath $\cal E$}}$ through the charges ${%
\mbox{\boldmath $\eta$}}_{ij}$ defined by relation 
\begin{equation}
{\mbox{\boldmath $\eta$}}_{ij}={\mbox{\boldmath $\epsilon$}}_{i}-{%
\mbox{\boldmath $\epsilon$}}_{j}.  \label{gcharge}
\end{equation}

According to our discussion from the previous Section, the production rates
of quarks and antiquarks in the chromoelectric field ${\mbox{\boldmath $\cal E$}}$ 
are 
\begin{equation}
\frac{dN^\pm_{i}}{d\Gamma } =  {\cal R}_{i}(\tau,p_\perp)
\delta \left( w \mp w_{i}\right) v,
\label{qrate}
\end{equation}
where we have defined
\begin{eqnarray}
 {\cal R}_{i}(\tau,p_\perp) &=&
\frac{\Lambda _{i}}{4\pi ^{3}}\left| \ln \left(
1-\exp \left( -\frac{\pi p_{\bot }^{2}}{\Lambda _{i}}\right) \right) \right|
\theta\left[2\tau\Lambda_i \sinh\left({\Delta \eta \over 4}\right)
- p_{\perp } \right], 
\label{Ri} \\ 
\nonumber \\
\nonumber \\
\Lambda _{i} &=& \left( g\left| {\mbox{\boldmath $\epsilon$}}_{i}\cdot {%
\mbox{\boldmath $\cal E$}}\right| -\sigma _{q}\right) \theta \left( g\left| {%
\mbox{\boldmath $\epsilon$}}_{i}\cdot {\mbox{\boldmath $\cal E$}}\right|
-\sigma _{q}\right),  
\label{Lami}
\end{eqnarray}
and 
\begin{equation}
w_{i}=-\frac{p_{\perp }^{2}}{2\Lambda _{i}}
\hbox{sign}\left( {\mbox{\boldmath $\epsilon$}}_{i}
\cdot {\mbox{\boldmath $\cal E$}}\right) .  
\label{wi}
\end{equation}

The quantity $\Lambda _{i}$ describes the effective force acting on the
tunneling quarks. The effect of the screening of the initial field by the
tunneling particles is taken into account by the subtraction of the
''elementary force'' characterized by the string tension $\sigma _{q}$.

Similarly, for gluons we have 
\begin{equation}
\frac{d\tilde{N}_{ij}}{d\Gamma } = 
\tilde{\mathcal{R}}_{ij}(\tau,p_\perp)
\delta \left( w-w_{ij}\right) v,
\label{grate}
\end{equation}
where  
\begin{eqnarray}
\tilde{\mathcal{R}}_{ij}(\tau,p_\perp) &=&
\frac{\Lambda _{ij}}{4\pi ^{3}}\left|
\ln \left( 1+\exp \left( -\frac{\pi p_{\bot }^{2}}{\Lambda _{ij}}\right)
\right) \right| 
\theta\left[2\tau\Lambda_{ij} \sinh\left({\Delta \eta \over 4}\right)
- p_{\perp } \right],
\label{Rij} \\
\nonumber \\
\nonumber \\
\Lambda _{ij} &=& \left(g\left| {\mbox{\boldmath $\eta$}}_{ij}\cdot {%
\mbox{\boldmath $\cal E$}}\right| -\sigma _{g}\right) \theta \left( g\left| {%
\mbox{\boldmath $\eta$}}_{ij}\cdot {\mbox{\boldmath $\cal E$}}\right|
-\sigma _{g}\right),
\label{Lamij}
\end{eqnarray}
and 
\begin{equation}
w_{ij}=-\frac{p_{\perp }^{2}}{2\Lambda _{ij}}
\hbox{sign}\left( {\mbox{\boldmath $\eta$}}_{ij}\cdot {%
\mbox{\boldmath
$\cal E$}}\right) .  
\label{wij}
\end{equation}
We note that the string tension of a tube spanned by gluons is three times
stronger than that of a quark tube, $\sigma _{g}=3\sigma _{q}$ \cite{DFcft}.

The implementation of the boost invariance in Eqs. (\ref{kineq}) and 
(\ref{kineg}) leads us to the following form of the transport equations 
\cite{BCDFosc} 

\newpage

\begin{equation}
\frac{\partial G^\pm_{i}}{\partial \tau }
\mp g{\mbox{\boldmath $\epsilon$}}_{i} \cdot 
\frac{d{\bf h}}{d\tau }\frac{\partial G^\pm_{i}}{\partial w}=\tau 
{\cal R}_{i}\left( \tau ,p_{\bot }\right) 
\delta \left( w \mp w_{i}(\tau, p_{\bot })\right) ,  \label{kineq1}
\end{equation}
\begin{equation}
\frac{\partial \tilde{G}_{ij}}{\partial \tau }
-g{\mbox{\boldmath $\eta$}}_{ij} \cdot 
\frac{d\mathbf{h}}{d\tau }\frac{\partial \tilde{G}_{ij}}{\partial w}=\tau 
\tilde{{\cal R}}_{ij}\left( \tau ,p_{\bot }\right)
\delta \left( w-w_{ij}(\tau ,p_{\bot })\right) .  \label{kineg1}
\end{equation}
Their formal solution is 
\begin{equation}
G^\pm_{i}\left( \tau ,w,p_{\bot }\right) =\int_{0}^{\tau }d\tau^{\prime }\ 
\tau^{\prime } \, {\cal R}_{i}\left(\tau^{\prime },p_{\bot }\right) 
\delta\left( \Delta h_{i}\left( \tau ,\tau ^{\prime }\right) 
\pm  w - w_i(\tau^\prime, p_{\bot }) \right),
\label{kineqs}
\end{equation}
\begin{equation}
\!\!\!\!\!
\tilde{G}_{ij}\left( \tau ,w,p_{\bot }\right) =\int_{0}^{\tau }d\tau
^{\prime }\ \tau ^{\prime }\ \tilde{{\cal R}}_{ij}\left( \tau
^{\prime },p_{\bot }\right) \delta \left( \Delta h_{ij}\left( \tau ,\tau
^{\prime }\right) +w - w_{ij}(\tau^\prime, p_{\bot })\right), \label{kinegs}
\end{equation}
where 
\begin{equation}
\Delta h_{i}\left(\tau ,\tau ^{\prime }\right) 
\equiv g{\mbox{\boldmath $\epsilon$}}_{i}\cdot 
\left[{\bf h}\left(\tau \right) -
{\bf h}\left(\tau ^{\prime }\right) \right], \nonumber 
\end{equation}
\begin{equation}
\Delta h_{ij}\left( \tau ,\tau^{\prime }\right) 
\equiv g{\mbox{\boldmath $\eta$}}_{ij} \cdot 
\left[{\bf h}\left( \tau \right) -
{\bf h}\left( \tau ^{\prime }\right) \right].
\label{hs}
\end{equation}
One may notice that the distribution functions (\ref{kineqs}) and (\ref{kinegs}) 
satisfy the following symmetry relations 
\begin{equation}
G^-_{i}\left( \tau ,w,p_{\bot }\right) =G^+_{i}\left( \tau
,-w,p_{\bot }\right) ,\quad \tilde{G}_{ij}\left( \tau ,w,p_{\bot
}\right) =\tilde{G}_{ji}\left( \tau ,-w,p_{\bot }\right) .
\label{symofg}
\end{equation}
We note also that the time integrals in (\ref{kineqs}) and (\ref{kinegs})
reveal the non-Markovian character of the particle production mechanism: the
behavior of the system at a time $\tau $ is determined by the whole
evolution of the system in the time interval $0\leq \tau ^{\prime }\leq 
\tau$.

\section{Color currents}

Equations (\ref{kineq}) and (\ref{kineg}) show how the particles behave under the
influence of the field. In order to obtain a self-consistent set of
equations we should have also the dynamic equation for the field. It can be
written in the following Maxwell form 
\begin{equation}
\partial _{\mu }{\bf F}^{\mu \nu }(x)={\bf j}^{\nu }(x)+{\bf j}%
_{D}^{\nu }(x),  \label{Maxwell}
\end{equation}
where ${\bf j}^{\nu }$ is the \textit{conductive current} (related to the
simple fact that particles carry color charges ${\mbox{\boldmath $\epsilon$}}%
_{i}$ and ${\mbox{\boldmath $\eta$}}_{ij})$ and ${\bf j}_{D}^{\nu }$ is
the \textit{displacement current} (induced by the tunneling of quarks and
gluons from the vacuum).

\subsection{\protect\bigskip Conductive current}

The form of the conductive current is standard 
\begin{eqnarray}
&& {\bf j}^{\nu }(x)=g\int dP \, p^{\nu }\left[ N_{f}\sum_{i=1}^{3}{%
\mbox{\boldmath
$\epsilon$}}_{i}\left( G^+_{i}(x,p)-G^-_{i}(x,p)\right)
+\sum_{i,j=1}^{3}{\mbox{\boldmath $\eta$}}_{ij}\tilde{G}_{ij}(x,p)\right].
\nonumber \\
&&
\label{condcur}
\end{eqnarray}
Substituting the quark and gluon distribution functions (\ref{kineqs}) and 
(\ref{kinegs}) into Eq. (\ref{condcur}), and using 
Eqs. (\ref{binvv3}) and (\ref{symofg}) we find  that ${\bf j}^\nu(x)$
has the following space-time structure 
\begin{equation}
{\bf j}^{\nu }(x)=\left[ {\bf j}^{0}(x),0,0,{\bf j}^{3}(x)\right]
=[z,0,0,t]{\mbox{\boldmath $\cal J$}}\left( \tau \right) ,  \label{condcur1}
\end{equation}
where
\begin{eqnarray}
{\mbox{\boldmath $\cal J$}}\left( \tau \right) &=&\frac{2g}{u}
\int d^{2}p_{\bot}  \frac{dw\ w}{v}
\left[ N_{f}\sum_{i=1}^{3}{\mbox{\boldmath $\epsilon$}}%
_{i}\,G^+_{i}\left( \tau ,w,p_{\bot }\right) +\sum_{i>j}^{3}{%
\mbox{\boldmath
$\eta$}}_{ij}\, \tilde{G}_{ij}\left( \tau ,w,p_{\bot }\right) \right].
\nonumber \\
& & 
\label{condcur2}
\end{eqnarray}
The use of the explicit form of the distribution functions $G_{i}$ and $%
\tilde{G}_{ij}$ in Eq. (\ref{condcur2}) gives 
\begin{eqnarray}
{\mbox{\boldmath $\cal J$}}\left( \tau \right)  &=&-\frac{2gN_{f}}{u}%
\sum_{i=1}^{3}{\mbox{\boldmath
$\epsilon$}}_{i}\int_{0}^{\tau }d\tau ^{\prime }\ \tau ^{\prime }\int
d^{2}p_{\bot }\frac{{\cal R}_{i}\left( \tau^\prime, p_{\bot }\right) \left[
\Delta h_{i}-w_{i}(p_{\perp },\tau ^{\prime })\right] }{\sqrt{\left[ \Delta
h_{i}-w_{i}(p_{\perp },\tau ^{\prime })\right] ^{2}+p_{\perp }^{2}u}}
\nonumber \\
&&-\frac{2g}{u}\sum_{i>j}^{3}{\mbox{\boldmath $\eta$}}%
_{ij}\int_{0}^{\tau }d\tau ^{\prime }\ \tau ^{\prime }\int d^{2}p_{\bot }%
\frac{\tilde{{\cal R}}_{ij}\left( \tau^\prime, p_{\bot }\right) 
\left[ \Delta
h_{ij}-w_{ij}(p_{\perp },\tau ^{\prime })\right] }{\sqrt{\left[ \Delta
h_{ij}-w_{ij}(p_{\perp },\tau ^{\prime })\right] ^{2}+p_{\perp }^{2}u}}.
\nonumber \\
&&
\label{condcur4}
\end{eqnarray}

\subsection{\protect\bigskip Displacement current}

The structure of the displacement current is less obvious. One can find the
form of ${\bf j}_{D}^{\nu }$ through the analysis of the energy and
momentum conservation laws for both the matter (quarks and gluons) and the
field 
\begin{equation}
\partial _{\mu }T_{\hbox{\small matter}}^{\mu \nu}(x)+
\partial _{\mu }T_{\hbox{\small field}}^{\mu \nu}(x)=0.
\label{enmomcon3}
\end{equation}
The $\nu =0$ component of this equation gives for the field part
\begin{equation}
\partial_{\mu }T_{\hbox{\small field}}^{\mu 0}
=\frac{\partial }{\partial t}\left( \frac{1}{2}
{\mbox{\boldmath $\cal E$}}^{2}\right) ={\mbox{\boldmath $\cal E$}\cdot }%
\frac{\partial {\mbox{\boldmath $\cal E$}}}{\partial t}=-{\bf F}%
^{30}\cdot \frac{\partial {\bf F}^{03}}{\partial t},  \label{enmomcon3a}
\end{equation}
and for the matter part
\begin{eqnarray}
\partial _{\mu }T_{\hbox{\small matter}}^{\mu 0} 
&=&-g{\bf F}^{\mu \nu }\cdot \int dP\ p^{0}\
p_{\nu }\ \partial _{\mu }^{p}\left[ N_{f}\sum_{i=1}^{3}{%
\mbox{\boldmath
$\epsilon$}}_{i}\left( G^+_{i}-G^-_{i}\right) +\sum_{i,j=1}^{3}{%
\mbox{\boldmath $\eta$}}_{ij}\tilde{G}_{ij}\right] \nonumber  \\
&&+\int dP\ p^{0}\left[ N_{f}\sum_{i=1}^{3}\left( \frac{dN^+_{i}}{d\Gamma }+%
\frac{d N^-_{i}}{d\Gamma }\right) +\sum_{i,j=1}^{3}\frac{d\tilde{%
N}_{ij}}{d\Gamma }\right] ,  \label{enmomcon3b}
\end{eqnarray}
where we used Eqs. (\ref{kineq}) and (\ref{kineg}). Using the last results we
may write 
\begin{eqnarray}
{\bf F}^{30} \cdot \frac{\partial {\bf F}^{03}}{\partial t} 
&=&{\bf F}^{30} \cdot g \int dP\, p^{3}
\left[ N_{f}\, \sum_{i=1}^{3}{\mbox{\boldmath
$\epsilon$}}_{i}\left( G^+_{i}-G^-_{i}\right) +\sum_{i,j=1}^{3}{%
\mbox{\boldmath $\eta$}}_{ij}\tilde{G}_{ij}\right]  
\nonumber \\
&&+{\bf F}^{30}\cdot \int dP\ \left[ N_{f}\ \sum_{i=1}^{3}\frac{p^{0}{%
\mbox{\boldmath $\epsilon$}}_{i}}{\ {\mbox{\boldmath $\epsilon$}}_{i}\cdot 
{\bf F}^{30}}\left( \frac{dN^+_{i}}{d\Gamma }+\frac{dN^-_{i}}{%
d\Gamma }\right) \right.  
\nonumber \\
&&+\left. \sum_{i>j}^{3}\frac{p^{0}{\mbox{\boldmath $\eta$}}_{ij}}{{%
\mbox{\boldmath
$\eta$}}_{ij}\cdot {\bf F}^{30}}\left( \frac{d\tilde{N}_{ij}}{d\Gamma 
}+\frac{d\tilde{N}_{ji}}{d\Gamma }\right) \right] .
\end{eqnarray}
Similarly, we may analyze the $\nu =3$ component of Eq. (\ref{enmomcon3}).
The conclusion is that the field equations (\ref{Maxwell}) represent a 
sufficient condition for the conservation of energy and momentum if the 
displacement current has the structure
\begin{equation}
{\bf j}_{D}^{\nu }(x)=\left[{\bf j}_{D}^{0}(x),0,0,{\bf j}%
_{D}^{3}(x)\right] =[z,0,0,t]{\mbox{\boldmath $\cal J$}}_{D}\left( \tau
\right)   \label{convcur1}
\end{equation}
where
\begin{eqnarray}
{\mbox{\boldmath $\cal J$}}_{D}\left( \tau \right)  &=&\frac{N_{f}}{\tau^2 }%
\sum_{i=1}^{3}\frac{{\mbox{\boldmath $\epsilon$}}_{i}}{{%
\mbox{\boldmath
$\epsilon$}}_{i}\cdot {\mbox{\boldmath $\cal E$}}}\int dw \,
d^{2}p_{\bot }\left( \frac{dN^+_{i}}{d\Gamma }+\frac{dN^-_{i}}{%
d\Gamma }\right)  \nonumber \\
&&\ \ \ +\frac{1}{\tau^2 }\sum_{i>j}^{3}\frac{{\mbox{\boldmath $\eta$}}_{ij}}{{%
\mbox{\boldmath
$\eta$}}_{ij}\cdot {\mbox{\boldmath $\cal E$}}}\int dw \,
d^{2}p_{\bot }\left( \frac{d\tilde{N}_{ij}}{d\Gamma }+\frac{d\tilde{N%
}_{ji}}{d\Gamma }\right) .  \label{convcur4}
\end{eqnarray}
Integration over $w$ in Eq. (\ref{convcur4}) can be easily done 
\begin{eqnarray}
{\mbox{\boldmath $\cal J$}}_{D}\left( \tau \right)  &=&
\frac{2  N_{f}}{\tau ^{2}}\sum_{i=1}^{3}\frac{{%
\mbox{\boldmath $\epsilon$}}_{i}}{{\mbox{\boldmath
$\epsilon$}}_{i}\cdot {\mbox{\boldmath $\cal E$}}}\int d^{2}p_{\bot }%
\sqrt{w_{i}^{2}+p_{\bot }^{2}u}  \,\,
{\cal R}_{i}\left(\tau, p_{\bot }\right)
\nonumber \\
&&\ \ +\frac{2}{\tau ^{2}}%
\sum_{i>j}^{3}\frac{{\mbox{\boldmath $\eta$}}_{ij}}{{\mbox{\boldmath
$\eta$}}_{ij}\cdot {\mbox{\boldmath $\cal E$}}}\int d^{2}p_{\bot }\sqrt{%
w_{ij}^{2}+p_{\bot }^{2}u} \,\,
\tilde{{\cal R}}_{ij}\left(\tau, p_{\bot }\right).
\label{convcur5}
\end{eqnarray}

We note that the form of Eqs. (\ref{condcur1}) and (\ref{convcur1})
implies that both the  conductive and displacement currents are 
conserved separately
\begin{equation}
\partial _{\nu }\,{\bf j}^\nu(x)=0, \hspace{1.5cm}
\partial _{\nu }\,{\bf j}_D^\nu(x)=0.
  \label{condcur3}
\end{equation}

We also note that the integral over $p_\perp$ is restricted by
condition (\ref{deltaeta}), implicitly included in 
${\cal R}_{i}$ and $\tilde{{\cal R}}_{ij}$. If we did not use
our finite-size correction, the displacement current would diverge
at small $\tau$, ${\mbox{\boldmath $\cal J$}}_{D}(\tau) \sim \tau^{-2}$, 
and the field equation would be singular, compare Eq. (\ref{biMaxwell0}).
In the standard case (with zero longitudinal momenta of the tunneling
particles and with no finite-size corrections), the quantities $w_i$ 
and $w_{ij}$ are zero, hence 
${\mbox{\boldmath $\cal J$}}_{D}(\tau) \sim \tau^{-1}$ for small $\tau$,
and the field equation is regular in the limit $\tau \rightarrow 0$

\subsection{Field equations}

With the all substitutions required by the boost invariance, the field
equation (\ref{Maxwell}) may be written as 
\begin{equation}
\frac{d{\mbox{\boldmath $\cal E$}}
\left( \tau \right) }{d\tau}=
-\tau\left[ {\mbox{\boldmath
$\cal J$}}\left( \tau \right) +{\mbox{\boldmath $\cal J$}}_{D}\left( \tau
\right) \right] \label{biMaxwell0}
\end{equation}
or
\begin{equation}
\frac{d^{2}\mathbf{h}\left( \tau \right) }{d\tau ^{2}}=\frac{1}{\tau }\frac{d%
\mathbf{h}\left( \tau \right) }{d\tau }+\tau ^{2}\left[ {\mbox{\boldmath
$\cal J$}}\left( \tau \right) +{\mbox{\boldmath $\cal J$}}_{D}\left( \tau
\right) \right] .  \label{biMaxwell}
\end{equation}
This is an integro-differential equation for the function $\mathbf{h}\left(
\tau \right)$, because the conductive current ${\mbox{\boldmath $\cal J$}}%
\left( \tau \right)$ depends not only on $\mathbf{h}\left( \tau \right) $
but also on all the values of $\mathbf{h(}\tau ^{\prime })$ for $0\leq \tau
^{\prime }\leq \tau $. Eq. (\ref{biMaxwell}) has to be solved numerically
step by step for given initial values. These are taken in the form 
\cite{BCDFosc,DFcft}
\begin{equation}
{\bf h}(0)=0,\quad \frac{1}{\tau }\frac{d{\bf h}}{d\tau }%
(0)=-{\mbox{\boldmath $\cal E$}}_{0}{\,=}-\sqrt{\frac{2\sigma _{g}}{\pi r^{2}%
}}k\mathbf{q}.  \label{initcon}
\end{equation}
Here the Gauss law has been used to determine the initial strength of
the chromoelectric field ${\mbox{\boldmath $\cal E$}}_{0}$ in terms of
the transverse radius of the color-flux-tube ($\pi r^2$= 1 fm$^2$),
the string tension ($\sigma_g= 3 \sigma_q =$ 3 GeV/fm), and the number
of color charges $k$.  Since the exchange of color charges at the
initial stage of a heavy-ion collision leads to the color fields
spanned by gluons \cite{bironk}, we assume that $\mathbf{q}$ is one of
the gluon color charges ${\mbox{\boldmath $\eta$}}_{ij}$. In practice
we take ${\bf q}={\mbox{\boldmath $\eta$}}_{12}$, so only the third
component of the chromoelectric field, ${\mbox{\boldmath $\cal
E$}}^{(3)}$, is present in our numerical calculations.  The solution
of Eq. (\ref{biMaxwell}) is independent of the initial condition for
$\mathbf{h}\left(\tau \right)$ because of the cancellations
connected with the gauge transformation which leaves ${\mbox{\boldmath
$\cal E$}}$ unchanged.

\begin{figure}[b]
\epsfysize=7cm
\par
\begin{center}
\mbox{\hspace{-4cm} \epsfbox{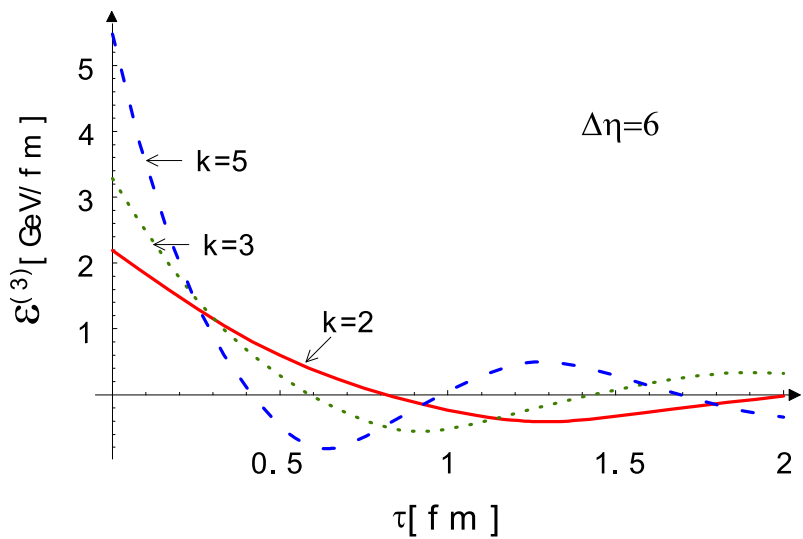}}
\end{center}
\caption{Time dependence of the chromoelectric field for different
values of $k$. The maximal quasirapidity interval allowed for tunneling
is fixed, $\Delta\eta$=6.}
\label{pole}
\end{figure}

\begin{figure}[b]
\epsfysize=7cm
\par
\begin{center}
\mbox{\hspace{-4cm} \epsfbox{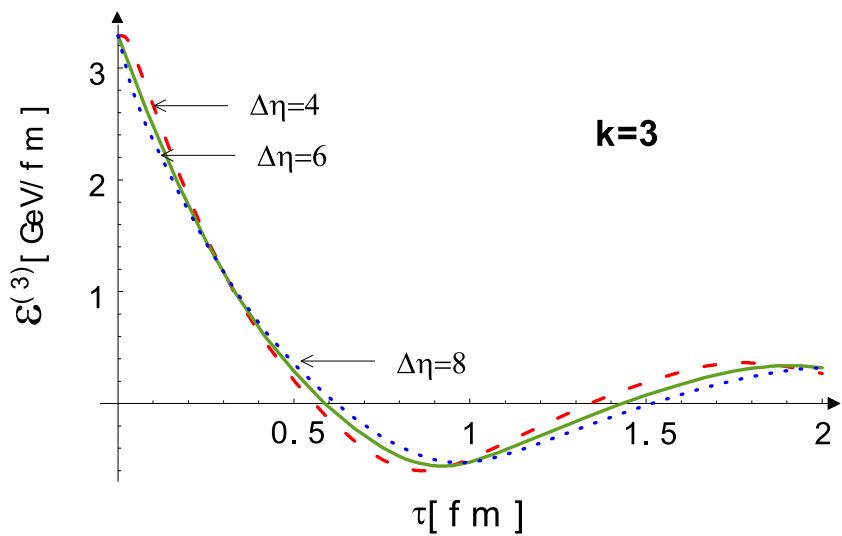}}
\end{center}
\caption{Time dependence of the chromoelectric field for different
values of $\Delta \eta$. The initial field strength is fixed,
$k=3$.}
\label{pole1}
\end{figure}

In Fig. \ref{pole} we plot the time dependence of the chromoelectric
field, as calculated from Eqs. (\ref{biMaxwell}) and (\ref{initcon})
for different values of the parameter $k$. The maximal allowed
quasirapidity interval $\Delta \eta$ is 6 in this case. We observe the
field oscillations with the frequency growing with $k$. The shape of
the oscillations is almost identical to those found before in
Ref. \cite{BCDFosc}. Clearly, the modification of the tunneling
process does not affect the field behavior in this case.  In
Fig. \ref{pole1} we show the time dependence of the field for the
fixed initial strength, $k=3$, and for different values of $\Delta
\eta $. In the three considered cases, $\Delta \eta$ = 4, 6 and 8, the
field oscillations are very similar. Looking in more detail, we
observe that with increasing $\Delta \eta $, the chromoelectric field
decreases faster at the very initial stage of the process, i.e., for
$0<\tau <0.1$ fm. Later the decrease of the field is weaker and for
$\tau \sim 0.3$ fm the values of the chromoelectric fields are
approximately the same for different values of $\Delta \eta$. For
longer times one may notice that the period of the field oscillations
is slightly longer for larger values of $\Delta \eta$ -- a faster
initial decay of the field causes a faster back reaction of the induced
currents, and the subsequent slow down of the decay.

\section{Energy density and pressure of the plasma}

The energy-momentum tensor of the quark-gluon plasma has a structure
\begin{equation}
T_{\hbox{\small matter}}^{\mu \nu }=\left[ \varepsilon \left( \tau \right)
+P\left( \tau \right) \right] u^{\mu }u^{\nu }-P\left( \tau \right) g^{\mu
\nu },  \label{tmunu}
\end{equation}
where in our two-dimensional model
\begin{equation}
u^{\mu }=\frac{1}{\tau }\left( t,z\right) .  \label{umu}
\end{equation}
It is important to emphasize that in our case the standard form of the
energy-momentum tensor, Eq. (\ref{tmunu}), does not follow from the assumption 
of the local thermodynamic equilibrium, but it is a direct consequence of the
boost-invariance. Eq. (\ref{tmunu}) can be derived directly from the definition 
of the 
energy-momentum tensor with the help of the symmetry relations (\ref{symofg}). 
The conservation law (\ref{enmomcon3}) together with Eq. (\ref{tmunu}) imply
\begin{equation}
\frac{d }{d \tau }\left[ \varepsilon \left( \tau \right)
+ \frac{1}{2}{\mbox{\boldmath $\cal E$}}^{2}(\tau)  \right] =-\frac{%
\varepsilon \left( \tau \right) +P\left( \tau \right) }{\tau }.
\label{bjork}
\end{equation}
In the absence of the fields Eq. (\ref{bjork}) is reduced to the Bjorken 
equation describing the evolution of the energy density in a boost-invariant
hydrodynamic model \cite{bjorken}. On the other hand, neglecting the expansion
effects, described by the term on the right-hand-side of Eq. (\ref{bjork}),
we obtain the simple conservation law for the total energy of the field 
and matter.

The non-equilibrium energy density $\varepsilon $ and the non-equilibrium
pressure $P$ are 
\begin{equation}
\varepsilon \left( \tau \right) =\frac{1}{u}\int d^{2}p_{\bot }\,\,dw\,v\,%
\left[ N_{f}\,\sum_{i=1}^{3}\left( G_{i}^{+}+G_{i}^{-}\right)
+\sum_{i,j=1}^{3}\tilde{G}_{ij}\right]   \label{epstau}
\end{equation}
and
\begin{equation}
P\left( \tau \right) =\frac{1}{u}\int d^{2}p_{\bot }\,\,dw\,\frac{w^{2}}{v}\,%
\left[ N_{f}\,\sum_{i=1}^{3}\left( G_{i}^{+}+G_{i}^{-}\right)
+\sum_{i,j=1}^{3}\tilde{G}_{ij}\right] .  \label{Peps}
\end{equation}

\begin{figure}[t]
\epsfysize=7cm
\par
\begin{center}
\mbox{\hspace{-4cm} \epsfbox{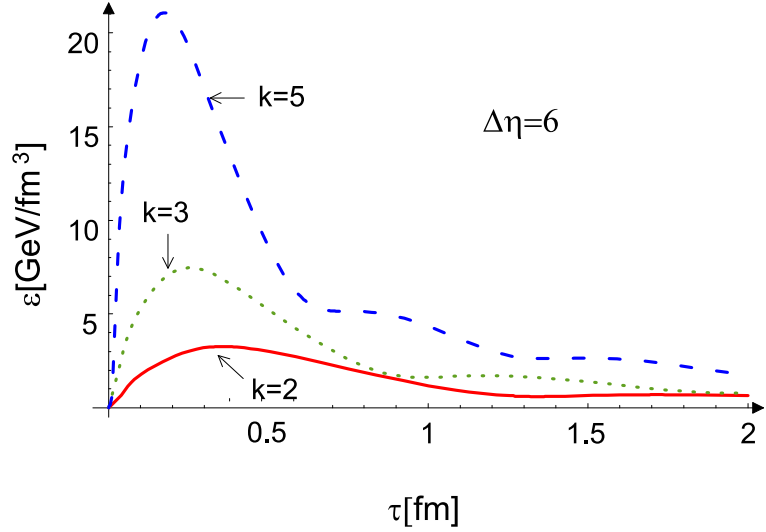}}
\end{center}
\caption{Time dependence of the energy density of the plasma, Eq. (57),
for different values of $k$ and $\Delta \eta$ =6. }
\label{ene}
\end{figure}

\begin{figure}[t]
\epsfysize=7cm
\par
\begin{center}
\mbox{\hspace{-4cm} \epsfbox{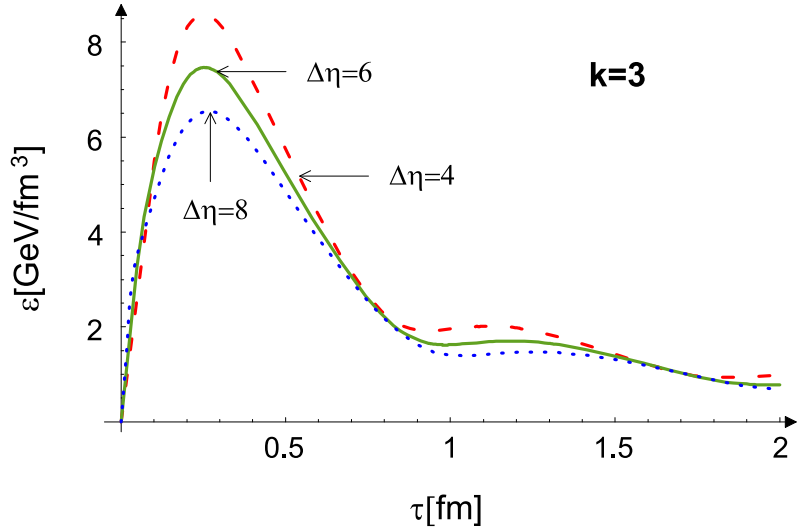}}
\end{center}
\caption{Time dependence of the energy density of the plasma
for different values of $\Delta \eta$ and $k=3$.}
\label{ene1}
\end{figure}

The time dependence of the energy density $\varepsilon \left( \tau \right)$,
following from Eq. (\ref{epstau}), is shown in Figs. \ref{ene} and 
\ref{ene1}. In Fig. \ref{ene} we fix the maximal quasirapidity interval 
$\Delta \eta =6$ and show the results for different strengths of the initial field, 
$k=$ 2, 3 and 5. We observe that the energy density grows very rapidly and reaches 
maximum at a fraction of a fermi. Later the energy density decreases, which 
is an effect connected with the longitudinal expansion of the system, imposed 
by the boost invariance. The maximal values of the energy density are very 
close to those found in Ref. \cite{BCDFosc}. In Fig. \ref{ene1} we show the 
results for the fixed initial value of the field, $k=3$, and for different 
quasirapidity intervals, $\Delta \eta =$ 4, 6 and 8. It is interesting
to observe that with  increasing values of  $\Delta \eta$ the maximal energy 
density gets smaller. This behavior is connected with the time dependence of 
the chromoelectric field. For larger values of  $\Delta \eta $ the decay of the
field is slower (except for the very beginning of the decay process) and the
growth of the energy density is weaker.

\begin{figure}[t]
\epsfysize=7cm
\par
\begin{center}
\mbox{\hspace{-4cm} \epsfbox{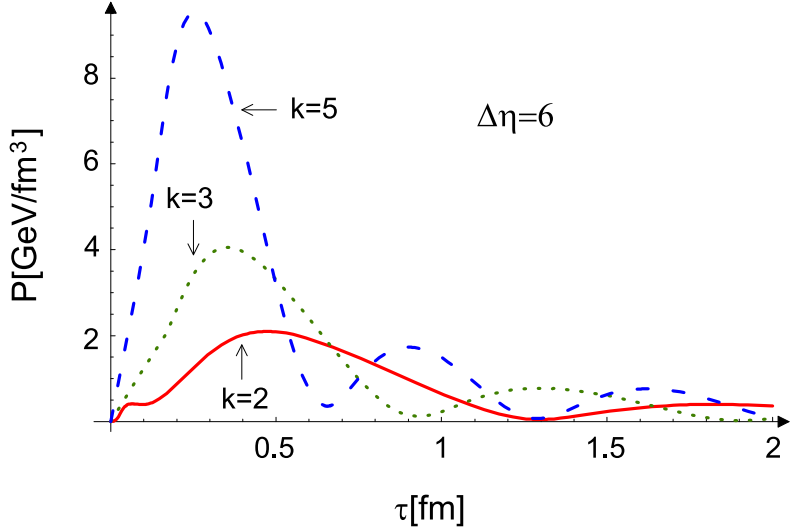}}
\end{center}
\caption{Time dependence of the pressure of the plasma, as defined by
Eq. (58), $k=2, 3, 5$ and $\Delta \eta$ = 6.}
\label{c}
\end{figure}

\begin{figure}[t]
\epsfysize=7cm
\par
\begin{center}
\mbox{\hspace{-4cm} \epsfbox{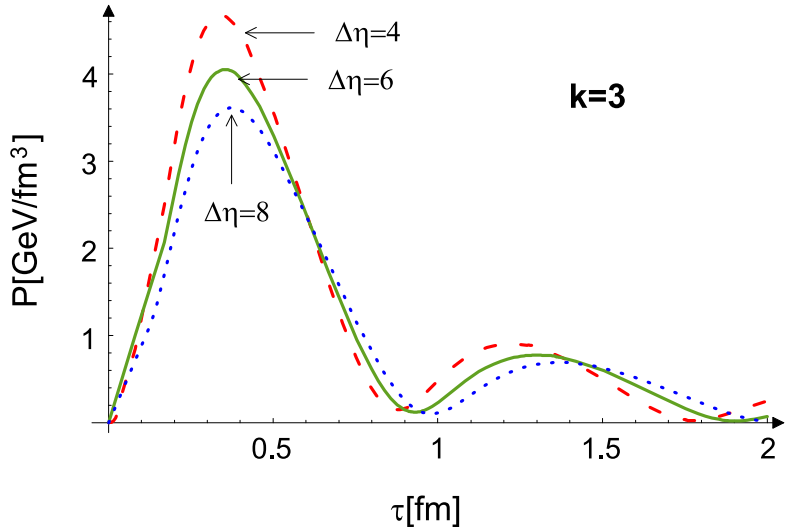}}
\end{center}
\caption{Time dependence of the pressure of the plasma, 
$k= 3$ and $\Delta \eta$ = 4, 6, 8.}
\label{c1}
\end{figure}

We have checked that our numerically evaluated functions $\varepsilon
(\tau)$ and $P(\tau)$ obey Eq. (\ref{bjork}).  The time-dependence of
the non-equilibrium pressure $P (\tau)$ is depicted in Figs.
\ref{c} and \ref{c1}. One can notice that the minima of $P(\tau)$ 
correspond to  the extremes of the chromoelectric field. At these points the 
longitudinal momenta of the particles practically vanish and the field is 
the strongest. This behavior reminds oscillations of a simple string. 
Another interesting feature of the process discussed here is that the ratio 
$P\left( \tau \right)/\varepsilon \left( \tau \right)$  oscillates
around the mean value $P/\varepsilon \sim 1/3$, which corresponds to the
equilibrium limit. Imposing condition
$P/\varepsilon = 1/3$ in Eq. (\ref{bjork}) we may treat this expression
as an equation determining $\varepsilon$ (the time dependent
chromoelectric field enters here as the only input). We have found
that the solutions of this equation are very good approximations of the
exact solutions shown in Fig. \ref{ene} and \ref{ene1}.

\section{Conclusions}

In this paper, the boost invariant tunneling has been incorporated into the
framework of the kinetic theory describing production of the quark-gluon
plasma in strong color fields. Our description of the tunneling process
includes the effect of a finite space-like distance formed between the
particles. Such a distance appears, since the energy and momentum
conservation laws should be satisfied locally during the tunneling. In our
approach, the particles emerging from the vacuum have finite longitudinal
momenta, which is a direct consequence of the tunneling along the
trajectories of constant invariant time.

We restrict the possibility of creation of pairs at very large distances (in
order to retain the boost-invariance of the system, this condition is
formulated in the quasirapidity space). In this way we mimic the
constraints imposed by the fact that total energies of the realistic systems
are finite. \ We find that the finite-size corrections are crucial to have
non-trivial solutions of the kinetic equations. For the realistic
finite-size corrections (as suggested by the accessible rapidity range in
the present experiments with the ultra-relativistic heavy-ions) we find the
solutions of the kinetic equations which are very close (qualitatively and
quantitatively) to the results of \ the previous investigations: the
chromoelectric fields oscillate, and large densities of quarks and gluons
are produced in a very short time. The increase of the region allowed for
tunneling does not lead to the increase of the \ produced maximal energy
density of quarks and gluons. This feature of the model can be used to
interpret small differences in many characteristics of the heavy-ion
collisions at the SPS and RHIC energies.

\end{document}